# The Landau-Lifshitz-Bloch equation for domain wall motion in antiferromagnets

Z. Y. Chen[1], Z. R. Yan[1], M. H. Qin[1,*], and J. –M. Liu[1,2]

[1]*Institute for Advanced Materials, South China Academy of Advanced Optoelectronics and Guangdong Provincial Key Laboratory of Quantum Engineering and Quantum Materials, South China Normal University, Guangzhou 510006, China*

[2]*Laboratory of Solid State Microstructures and Innovative Center for Advanced Microstructures, Nanjing University, Nanjing 210093, China*

**[Abstract]** In this work, we derive the Landau-Lifshitz-Bloch equation accounting for the multi-domain antiferromagnetic (AFM) lattice at finite temperature, in order to investigate the domain wall (DW) motion, the core issue for AFM spintronics. The continuity equation of the staggered magnetization is obtained using the continuum approximation, allowing an analytical calculation on the domain wall dynamics. The influence of temperature on the static domain wall profile is investigated, and the analytical calculations reproduce well earlier numerical results on temperature gradient driven saturation velocity of the AFM domain wall, confirming the validity of this theory. Moreover, it is worth noting that this theory could be also applied to dynamics of various wall motions in an AFM system. The present theory represents a comprehensive approach to the domain wall dynamics in AFM materials, a crucial step toward the development of AFM spintronics.

*qinmh@scnu.edu.cn

As promising materials for spintronics, antiferromagnets have attracted significant attention recently because they show fast magnetic dynamics and produce non-perturbing stray fields,[1-4] especially after the effective detection and manipulation of antiferromagnetic (AFM) state were experimentally realized.[5-7] Theoretically, the spin dynamics in an AFM lattice can also be investigated using the Landau-Lifshitz-Gilbert (LLG) equation based on the atomistic spin models, and a number of driving mechanisms[8-16] have been proposed to drive effectively the domain wall (DW) in an AFM lattice. These important works not only contribute a lot to fundamental physics but also do provide useful information for potential AFM spintronic devices.

Nevertheless, for a realistic spintronic device where the lattice size under consideration is huge, atomistic spin models are far from sufficient and an efficient computation based on such atomistic models becomes non-realistic due to the computation capacity limit. As a good approximation, one can utilize the coarse-grain scheme and use a macro-moment **m** to express the magnetization of a finite region (called as a grain) inside a ferromagnetic (FM) domain, and thus the LLG equation on the macro-moment **m** can be used without increasing much the computational cost. For a multi-domain FM lattice, each domain can be divided into a number of such regions each with its own macro-moment, and the dynamics of domain evolution and domain wall motion can thus be tracked efficiently using the LLG equation.

Unfortunately, the high efficient computation based on the LLG equation becomes challenging for an AFM system where no such coarse-grain scheme can be utilized, since an AFM lattice consists of two inter-crossing and spin-antiparallel sublattices. Consider that an AFM domain wall may have spatial width as large as ~ 10 nm, one sees that the whole lattice used for the LLG-based micromagnetic simulation must be at least as large as ~100 nm if wall motion is considered. This makes a computation impossible due to the capacity limit, unless the lattice is cut down to ~ 10 nm. At a cost of physical reality, one has to set the axial anisotropies two orders of magnitude stronger than realistic values, and the domain wall becomes unreasonably narrow (~ 1 nm). This is the first major issue to handle. On the other hand, white noise terms are usually included into the effective field for the LLG dynamics in order to simulate temperature ($T$)-dependent effects, which also add huge computation cost to the simulations. Moreover, it is known that the magnetization magnitude is a function of $T$, usually it decreases with increasing $T$ until the transition point $T_c$ and this fact is hardly described by the LLG-based simulations, as pointed out in earlier work. This is the second major issue. It should be mentioned that other theoretical approaches such as the Lagrangian scheme seem either hard to describe reasonably the AFM wall dynamics at finite $T$. Thus, realistic appealing to numerical approaches is thus raised in order to treat the wall dynamics in an AFM system at finite $T$ or $T$-gradient, noting that the $T$-relevant controls, e.g. $T$-gradient driven wall motions have been often taken in the AFM spintronic devices.

In short, there is an urgent need to develop an approach in dealing with continuum models for AFM lattice so that computation load can be reduced. Along this line, it is noted that wall motion in a FM lattice under a $T$-gradient field has been simulated using the Landau-Lifshitz-Bloch (LLB) equation.[17-18] This computation has been proven to be efficient in large-scale micromagnetic simulation of realistic spintronic devices at high $T$ and in short time. Reasonable results on the wall motion and Walker breakdown in a multi-domain FM lattice have been obtained within the framework of the LLB equation. Most recently, the results on the multi-domain FM lattice suggested a linear relation between the wall velocity and $T$-gradient. This relation was once applied to describe the domain wall motion in an AFM lattice. Unfortunately, this relation agrees with numerical results under small $T$-gradient, but deviates seriously when the $T$-gradient is large.[14] It is either inconsistent with the fact that the wall velocity should be limited by the maximum spin-wave group velocity.[9-10] It would thus be highly concerned and interested to ask if LLB scheme can be applied to track efficiently the domain wall motion in a large and multi-domain AFM lattice. Indeed, the LLB equation on a ferrimagnetic (FiM) mono-domain lattice was recently proposed,[19-20] which becomes the basis for deriving a generalized equation for a multi-domain AFM lattice.

In this Letter, we perform a derivation of the LLB equation for a multi-domain AFM lattice at finite $T$ and this equation would be highly efficient for large-scale micromagnetic simulation of realistic AFM spintronic devices. More importantly, a continuity equation for the staggered magnetizations can be derived out from this equation using the continuum approximation, which allows an analytical calculation on the domain wall motion in an AFM lattice (e.g. driven by a finite $T$-gradient). It is found that the theory's predictions about several crucial issues agree well with numerical results in literature.

We start from an AFM lattice with two inter-crossing FM sublattices whose spin alignments are antiparallel. We apply the coarse-grain scheme to the whole lattice divided into a number of grains as shown in Fig. 1. The grain size should be sufficiently large for high efficiency computation but sufficiently small in comparison with the concerned characteristic scales in lattice, e.g. domain wall width or other anomalies in the present case. The basic strategy is to track the magnetization evolution of the two sublattices separately. For an arbitrary grain (*i*) containing two FM sublattices (*v*, *κ*), if no interaction of this grain with its neighbors is considered, the LLB equation for magnetization $\mathbf{m}_v$ of sublattice *v* is written as:[19]

$$\frac{1}{\gamma_v}\frac{d\mathbf{m}_v}{dt} = \mathbf{m}_v \times \mathbf{H}_v + \alpha_\parallel \frac{\mathbf{m}_v \cdot \mathbf{H}_v}{m_v^2}\mathbf{m}_v - \alpha_\perp \frac{\mathbf{m}_v \times (\mathbf{m}_v \times \mathbf{H}_v)}{m_v^2}, \quad (1)$$

where $\gamma_v$ is the gyromagnetic ratio, $\alpha_\parallel/\alpha_\perp$ is the *T*-dependent longitudinal / transverse damping constants, $\mathbf{H}_v = \mathbf{H} + \mathbf{H}_{A,v} + \mathbf{H}_{v\kappa}$ is the effective field including external field $\mathbf{H}$, anisotropy field $\mathbf{H}_{A,v}$ and internal exchange field $\mathbf{H}_{v\kappa}$, assuming the *z*-axis as the easy axis. The internal exchange field $\mathbf{H}_{v\kappa}$ accounts the interaction between the sublattices *v* and *κ*. They are respectively given by:[19]

$$\mathbf{H}_{A,v} = -\frac{1}{\tilde{\chi}_{v,\perp}}(m_{x,v}\mathbf{e}_x + m_{y,v}\mathbf{e}_y), \quad \mathbf{m}_v = (m_{x,v}\mathbf{e}_x, m_{y,v}\mathbf{e}_y, m_{z,v}\mathbf{e}_z), \quad (2)$$

and

$$\mathbf{H}_{v\kappa} = -\frac{J_{0,v\kappa}}{\mu_v}\frac{\mathbf{m}_v \times (\mathbf{m}_v \times \mathbf{m}_\kappa)}{m_v^2} - \frac{1}{2}\left[\frac{1}{\Lambda_{vv}}\left(\frac{m_v^2}{m_{e,v}^2} - 1\right) - \frac{1}{\Lambda_{v\kappa}}\left(\frac{\tau_\kappa^2}{\tau_{e,\kappa}^2} - 1\right)\right]\mathbf{m}_v, \quad (3)$$

where detailed definitions of transverse susceptibility $\chi_{v,\perp}$, coupling constant $J_{0,v\kappa}$, saturation moment $\mu_v$, equilibrium magnetization $m_{e,v}$, longitudinal rates $\Lambda_{vv}$ and $\Lambda_{v\kappa}$, and intermediate variables $\tau_\kappa$ and $\tau_{e,\kappa}$ can be found in Ref. 19. The first and third terms in the right side of Eq. (1) have the same forms as those in the LLG equation, and the second term describes the longitudinal relaxation depicting the magnitude variation of magnetization due to thermal fluctuations at finite *T*.

It is noted that the *T*-dependent parameters in the two sublattices equal each other (e.g. $\gamma_v = \gamma_\kappa = \gamma$, $m_{e,v} = m_{e,\kappa} = m_e$, $\mu_v = \mu_\kappa = \mu_S$, $\chi_{v,\perp} = \chi_{\kappa,\perp} = \chi_\perp$), and $\mathbf{H}_{v\kappa}$ has a more compact form:

$$\mathbf{H}_{\nu\kappa} = -\frac{J_0}{\mu_S}\frac{\mathbf{m}_\nu \times (\mathbf{m}_\nu \times \mathbf{m}_\kappa)}{m_\nu^2} - \frac{1}{2}\left[\frac{1}{\tilde{\chi}_\parallel}\left(\frac{m_\nu^2}{m_e^2} - 1\right) + \frac{|J_0|}{\mu_S}\frac{m_\nu^2 - \tau_\kappa^2}{m_e^2}\right]\mathbf{m}_\nu, \qquad (4)$$

where $\chi_\parallel$ is the longitudinal susceptibility, $J_{0,\nu\kappa} = J_0 = N_D J$ with $J$ is the exchange coupling between the nearest neighbor atomistic spins and $N_D$ is the coordination number. Following the earlier works,[21-23] these parameters $m_e$, $\chi_\parallel$, and $\chi_\perp$, are reasonably estimated by numerical simulations using the stochastic LLG equation based on the atomistic model. As an example, we present the estimated parameters (empty points) and corresponding fitted results (solid lines) in Fig. 2, given the uniaxial anisotropy 0.02$J$. Their good consistencies confirm the estimations.

Subsequently, we discuss the effect of $T$. It is noted that thermal fluctuations are less dependent on spin structures, and thus the stochastic fields for a FM system can be approximately applicable to an AFM system.[24-26] This argument has been confirmed in earlier work which calculates the stochastic fields strictly using the Fokker-Planck equation.[23] When the stochastic fields are considered, the LLB equation for grain ($i$) now reads:

$$\frac{1}{\gamma}\frac{d\mathbf{m}_\nu}{dt} = \mathbf{m}_\nu \times \mathbf{H}_\nu + \alpha_\parallel \frac{\mathbf{m}_\nu \cdot \mathbf{H}_\nu}{m_\nu^2}\mathbf{m}_\nu - \alpha_\perp \frac{\mathbf{m}_\nu \times [\mathbf{m}_\nu \times (\mathbf{H}_\nu + \boldsymbol{\xi}_{\perp,\nu})]}{m_\nu^2} + \boldsymbol{\xi}_{\parallel,\nu}, \qquad (5)$$

where $\boldsymbol{\xi}_{\parallel,\nu}/\boldsymbol{\xi}_{\perp,\nu}$ is the longitudinal / transverse stochastic field with

$$\langle \xi_{\eta,\nu}^a(t,\mathbf{r})\xi_{\eta,\nu}^b(t',\mathbf{r}')\rangle = 2D_\eta \delta_{ab}\delta(t-t')\delta(\mathbf{r}-\mathbf{r}'), \quad \eta = (\parallel, \perp), \qquad (6)$$

where $a$, $b$ are the Cartesian components ($= x, y, z$), and the longitudinal and transverse diffusion constants $D_\parallel$ and $D_\perp$ read respectively

$$D_\parallel = \frac{\alpha_\parallel \gamma k_B T}{M_S V}, \text{ and } D_\perp = \frac{(\alpha_\perp - \alpha_\parallel)k_B T}{\gamma M_S V \alpha_\perp^2}, \qquad (7)$$

where $k_B$ the Boltzmann constant, $M_S$ the saturation magnetization, and $V$ the grain volume.

Actually, any grain must have coupling with its neighbors and an inclusion of the coupling is a prerequisite to consider a multi-domain AFM system. We discuss the inter-grain exchange field between grain ($i$) and grain ($j$), using the same approach as given in Ref. 22 to extend the LLB equation. For two neighboring grains ($i$) and ($j$), the inter-grain exchange interaction $H_{exij}$ reads

$$H_{exij} = -J\sum_{\langle k,l \rangle} \mathbf{S}_k \cdot \mathbf{S}_l = -J\frac{F}{2a_l^2}\left(\frac{\mathbf{m}_{v,i}}{m_{v,i}} \cdot \frac{\mathbf{m}_{\kappa,j}}{m_{\kappa,j}} + \frac{\mathbf{m}_{\kappa,i}}{m_{\kappa,i}} \cdot \frac{\mathbf{m}_{v,j}}{m_{v,j}}\right), \tag{8}$$

where $\langle k, l \rangle$ sums all the nearest neighbor pairs connecting the two grains, $S$ is the normalized atomistic spin, $F$ is the interface area, and $a_l$ is the lattice constant, $\mathbf{m}_{v,i}$ / $\mathbf{m}_{\kappa,j}$ is the magnetization of sublattice $v$ / $\kappa$ in grain $i$ / $j$. Then, we obtain the inter-grain exchange field to sublattice $v$ of grain ($i$) imposed by sublattice $\kappa$ in grain ($j$)

$$\mathbf{H}_{ex,v,i} = -\frac{1}{M_S V/2}\frac{\partial H_{exij}}{\partial \mathbf{m}_{v,i}} = \frac{2A(0)}{a_l d M_S m_e^2}\left(\mathbf{m}_{\kappa,j} + \mathbf{m}_{v,i}\right), \tag{9}$$

where $A(0) = J/2a_l$ is the exchange stiffness at zero $T$, and $d$ is the grain dimension. It is noted that Eq. (9) is obtained on the assumption that the two sublattices' magnetizations in grains ($i$) and ($j$) can be described as macro-spins $\mathbf{m}_{v,i}$ and $\mathbf{m}_{\kappa,j}$. This would overestimate the inter-grain exchange coupling. Following the earlier work, a correction factor $a_l/d$ should be taken into account to diminish the overestimation.[22]

Moreover, considering the thermal fluctuations, the exchange stiffness is also $T$-dependent, given by $A(T) = A(0)m_e^2$ if the thermal average spin moment is equal to the equilibrium magnetization $m_e$. Thus, the total inter-grain exchange field of sublattice $v$ in grain $i$ reads

$$\mathbf{H}_{ex,v,i} = \frac{2A(T)}{d^2 M_S m_e^2}\sum_j \left(\mathbf{m}_{\kappa,j} + \mathbf{m}_{v,i}\right), \tag{10}$$

where the sum is over all the nearest neighboring grains.

To this stage, we have successfully obtained the LLB equation applicable to a multi-domain AFM lattice, in particular, to describe the domain wall dynamics. Certainly, a more explicit form of the LLB equation using the continuum approximation would be appreciared.[16] In proceeding, we define the total magnetization $\mathbf{m}_i = \mathbf{m}_{v,i} + \mathbf{m}_{\kappa,i}$ and staggered magnetization $\mathbf{n}_i = \mathbf{m}_{v,i} - \mathbf{m}_{\kappa,i}$ for grain ($i$) to replace $\mathbf{m}_{v,i}$ and $\mathbf{m}_{\kappa,i}$. The effective fields for grains ($i$) and ($j$) are then written as $\mathbf{H}_{v,i} = \mathbf{H}_{m,i} + \mathbf{H}_{n,i}$, and $\mathbf{H}_{\kappa,i} = \mathbf{H}_{m,i} + \mathbf{H}_{n,i}$, where $\mathbf{H}_{m,i}$ and $\mathbf{H}_{n,i}$ are respectively the effective fields related to $\mathbf{m}_i$ and $\mathbf{n}_i$. Noting that the longitudinal relaxation of sublattice magnetization is much faster than the transverse relaxation, and the magnetization is nearly identical to the equilibrium one, i. e., $|\mathbf{m}_{v,i}| = m_{e,i}$,[18-19] one has the alternative expressions of the LLB equations after necessary substitutions and continuum

approximation:

$$\frac{d\mathbf{m}}{dt} = \gamma(\mathbf{m}\times\mathbf{H}_m + \mathbf{n}\times\mathbf{H}_n) - \frac{\alpha_\perp}{2m_e^2}\left(\mathbf{m}\times\frac{d\mathbf{m}}{dt} + \mathbf{n}\times\frac{d\mathbf{n}}{dt}\right)$$
$$+\frac{\gamma\alpha_\|}{2m_e^2}\left[(\mathbf{m}\cdot\mathbf{H}_n)\mathbf{n} + (\mathbf{n}\cdot\mathbf{H}_m)\mathbf{n}\right]$$ (11)

and

$$\frac{d\mathbf{n}}{dt} = \gamma\mathbf{n}\times\mathbf{H}_m,$$ (12)

with the effective fields $\mathbf{H}_m$ and $\mathbf{H}_n$ (see Supplementary Material A for detailed derivation). Here, Eq. (11) has been transformed into the Gilbert form, and particular damping terms are safely omitted as done in the LLG scheme,[16,27-28] which hardly affects our main results.

For an AFM system below its Néel temperature ($T_N$), one has $\mathbf{m}\cdot\mathbf{n}\sim 0$, and $\mathbf{n}^2\sim 4m_e^2$. Under zero applied field, $\mathbf{m}$ as a function of $\mathbf{n}$ can be derived from Eq. (12):[16,28]

$$\mathbf{m} = \frac{\frac{d\mathbf{n}}{dt}\times\mathbf{n}}{4\gamma m_e^2(J_0/\mu_S + 2N_D A/d^2 M_S m_e^2)} = A_m\frac{d\mathbf{n}}{dt}\times\mathbf{n},$$ (13)

where parameter $A_m$ is introduced for brevity. Substituting Eq. (13) into Eq. (11) and taking the cross product with $\mathbf{n}$, we obtain:

$$A_m\mathbf{n}\times\frac{d^2\mathbf{n}}{dt^2}\times\mathbf{n} = \mathbf{n}\times\left(-\frac{\gamma}{2\tilde{\chi}_\perp}n_z e_z + \frac{\gamma A(T)}{M_S m_e^2}\nabla^2\mathbf{n} + \frac{\alpha_\perp}{2m_e^2}\frac{d\mathbf{n}}{dt}\right)\times\mathbf{n},$$ (14)

where $n_z$ is the z component of $\mathbf{n}$. It should be mentioned that this is the first time to obtain an analytical expression of the staggered magnetization for an AFM lattice, whose magnitude and orientation are spatially inhomogeneous and T-dependent. It thus allows one to track various stimuli-driven domain structure evolution and wall motion in a multi-domain AFM system.

By using Eqs. (13) and (14), we can perform the analytical calculations within the framework of the LLB equation for an AFM system. Here, the second-order derivative of $\mathbf{n}$ with respect to time is essential in distinguishing the magnetic dynamics in an AFM system from that in a FM one.[32] In particular, the parameters and magnitude for the staggered magnetization ($\mathbf{n}$) are T-dependent, allowing one to investigate the magnetic dynamics at finite T, including the domain wall motion in ultra-large scale. Furthermore, the domain wall motion in an AFM lattice, as driven by various stimuli such as temperature gradient,[13-15] external field,[27,29-30] and Néel spin-orbit torque,[10,31] can be similarly calculated using Eq. (14).

For the validity of this continuum LLB theory on AFM lattice, one consults to several well known facts for checking. As an initial check, we discuss the static solutions. One of the special solutions to Eq. (14) is the static Néel wall configuration with the polar angle of the staggered magnetization $\theta = 2\arctan[\exp(z - z_0)/\lambda]$, where $z_0$ is the position of the wall center, and $\lambda$ is the $T$-dependent wall width:

$$\lambda(T) = \sqrt{\frac{2\tilde{\chi}_\perp |A(T)|}{M_S m_e^2}}, \tag{15}$$

One observes that $\lambda(0)$ is exactly the same as that derived from the LLG equation.[16] Moreover, $\lambda$ increases with increasing $T$ and ultimately becomes divergent at $T_N$, as shown in Fig. 3(a) which gives the numerical and analytical calculated $\lambda$ as a function of $T$. The analytical data well coincide with the numerical results, both based on the LLB equation, supporting the validity of this continuum theory.

It is noted that the AFM domain wall profiles have important influence on the wall dynamics and relevant magnetoresistance, while their $T$-dependences are still unclear so far. We numerically study the effect of temperature on the Bloch domain wall profiles using Eq. (5) on an $8a_l \times 8a_l \times 200a_l$ system. Similar to ferromagnets,[33-35] three types of walls including circular, elliptical and linear walls are observed. The circular wall emerges at zero $T$, as shown in Fig. 3 (b) which gives the three components of the magnetization versus $y$ coordinate. Fig. 3(c) presents the components at $T = 1.4\ J/k_B$, which clearly demonstrates an elliptical wall. Similarly, the wall profiles can be described by the hyperbolic functions $n_z(T) = h_z(T)\tanh[(y - y_0)/\lambda(T)]$ and $n_t(T) = h_t(T)\mathrm{sech}[(y - y_0)/\lambda(T)]$, where $n_t$ is the transverse component of **n**, and $h_z/h_t$ is the amplitude of easy axis/transverse magnetization. The estimated $h_z(T)$ and $h_t(T)$ are summarized in Fig. 3(d) where $h_t$ is smaller than $h_z$ at finite $T$, demonstrating the existence of elliptical walls. In additions, for $T_h < T < T_N$, the domain wall is linear with a finite $h_z$ and zero $h_t$. This effect can be understood from the influence of thermal fluctuations on the domain wall. The spins in the wall usually deviate from the easy axis and have large exchange and anisotropy energies, and thus they are more sensitive to thermal fluctuations than the spins inside the domain, resulting in the fact that $h_t$ decreases more quickly than $h_z$ as $T$ increases, as confirmed in our simulations. Furthermore, the difference between $h_z(T)$ and $h_t(T)$ increases with the increasing anisotropy (the corresponding results are not shown here), the same as in FM systems.[33-34] As a matter of fact, earlier work claimed that the FM and AFM domain walls share common static properties at zero $T$.[36] Here, it is clearly demonstrated that this behavior also exists at finite $T$ even near $T_N$.

Given the validity of the developed LLB theory, we intend to solve Eq. (14) using the approach proposed in earlier work[37] to investigate the thermally driven domain wall motion for an AFM lattice in a finite $T$-gradient. For simplicity, we assume that $T$ is rather below $T_N$ and the domain wall structure is robust during its motion.[9,37-39] In this case, the staggered magnetization is a function of the composite variable $Z = z - vt$:

$$\frac{dn_x}{dt} = -vn'_x, \quad \frac{dn_z}{dt} = -vn'_z, \quad \frac{d\theta}{dt} = -v\theta' = -v\frac{\sin\theta}{\lambda}, \tag{16}$$

where $v$ is the wall velocity, and $n'/\theta'$ represents the derivative of $n/\theta$ with respect to $z$. Considering that the exchange interaction is much stronger than the magnetic anisotropy, we obtain the velocity of wall motion under a temperature gradient:[37]

$$v = \frac{-\frac{1}{\alpha_1} + \sqrt{\frac{1}{\alpha_1^2} + 4\alpha_2}}{2\alpha_2}, \tag{17}$$

where $\alpha_1$ and $\alpha_2$ are $T$-dependent variables (see Supplementary Material B for details). More interestingly, it is clearly shown that even for a very large $T$ gradient, the domain wall velocity will be limited by the saturation value:

$$v_{max} = \frac{\gamma a_l J \sqrt{2N_D}}{\mu_S} m_e = c(T), \tag{18}$$

where $c(0)$ is exactly the group velocity of spin wave at zero $T$.[9,16] Thus, it is confirmed again that the limitation of the domain wall velocity originates from the emission of spin wave under large driving fields. Furthermore, the propagation velocity of spin wave $c(T)$ is dependent of the strength of exchange interaction. With increasing $T$, the enhanced thermal fluctuations effectively weaken the interaction and in turn decreases $c(T)$ and the saturation domain wall velocity $v_{max}$. Thus, this constraint is considered and $\alpha_2$ is reasonably modified to estimate the domain wall velocity more precisely, as explicitly explained in the

Supplementary Material C. In Fig. 4, we compare the calculated DW velocity with the numerical result reproduced from Ref. 14 to further check the validity of our derivation. The two results are well in consistent with each other even in the whole investigated range of the $T$ gradient, strongly confirming our analytical calculations.

At last, we use another approach with polar coordinates to solve Eq. (14) to study the effect of the magnetic anisotropy and obtain

$$v = \frac{-\frac{1}{\alpha_{K,1}} + \sqrt{\frac{1}{\alpha_{K,1}^2} + 4\alpha_{K,2}}}{2\alpha_{K,2}}, \tag{19}$$

where $\alpha_{K,1}$ and $\alpha_{K,2}$ are related to the magnetic anisotropy (see Supplementary Material D for more details). As pointed out in the earlier work, AFM skyrmions can be driven efficiently by the magnetic anisotropy gradient.[40] This driving method also works for AFM domain wall, which has been confirmed in our unpublished work. Noting that the effective anisotropy is also dependent on $T$ (see $\chi_\perp(T)$ curve in Fig. 2(b)), the $T$ gradient could induce a magnetic anisotropy gradient which affects the wall motion as well. Certainly, this effect is not so significant for small $T$ gradients that Eq. (19) could be approximated replaced by Eq. (17).

Clearly, the validity of the dynamic equation for staggered magnetization in an AFM lattice has been well confirmed by checking the static domain wall profiles and $T$-gradient driven wall motion which are well consistent with earlier numerical results. Thus, the two major issues (AFM wall motion at finite $T$ in large scale system) which are hardly reached in the LLG-based simulations have been well removed if the LLB equation and derived continuum equation are utilized. More importantly, we would like to point out that this essential equation can be also used to investigate the AFM dynamics driven by other stimulis.[41] For example, a large-scale system is needed to generate Gauss $T$ field, which is hardly reached by the conventional LLG simulations. As a matter of fact, the analytical calculation has been performed, and the corresponding results will be reported elsewhere.

In conclusion, we have derived the LLB equation with inter-grain and stochastic fields for AFM systems, which allows one to investigate the magnetic dynamics at finite temperatures using multi-scale approaches. Moreover, the continuity equation of the staggered magnetization has been also derived using the continuum approximation. The derivations have been used to investigate the influence of temperature on the static AFM domain wall, which reveals a similar behavior to FM systems. More interestingly, the analytical calculation of the temperature gradient driven AFM domain wall motion well agrees with the numerical results and reproduces successfully the saturation velocity, well confirming the validity of our derivations. Importantly, this theory could be applied to other wall driving mechanisms such as Néel spin-orbit torques and spin transfer torques as well.

**Supplementary Materials**

A. Continuum equation.
B. The AFM domain wall velocity under a temperature gradient.
C. The derivation and modification of group velocity of spin waves.
D. The effect of magnetic anisotropy gradient induced by temperature gradient.

**Acknowledgment**

The work is supported by the National Key Projects for Basic Research of China (Grant No. 2015CB921202), and the Natural Science Foundation of China (No. 11204091), and the Science and Technology Planning Project of Guangdong Province (Grant No. 2015B090927006), and the Natural Science Foundation of Guangdong Province (Grant No. 2016A030308019).

**FIGURE CAPTIONS**

Fig.1. (color online) (top) Spin configuration of atomistic regular AFM lattice, where the whole AFM lattice is divided into lots of grains (regions). (bottom) Sublattice magnetization in a grain is described by two antiparallel macro-spins $\mathbf{m}_\nu$ and $\mathbf{m}_\kappa$.

Fig.2. (color online) The stochastic LLG simulated (a) $m_e$ and (b) $\chi_\parallel$ and $\chi_\perp$ as functions of temperature and the corresponding fitting results.

Fig.3. (color online) (a) The numerical and analytical calculated $\lambda$ as a function of $T$, and the three components of the magnetization versus $y$ coordinate at (b) $T = 0$, and (c) $T = 1.4$, and (d) the estimated $h_z$ and $h_t$ as functions of $T$. The sketches of circular and elliptical DWs are also respectively presented in the inserts of (b) and (c).

Fig.4. (color online) Numerical (empty circles) and analytical (solid line) calculated DW velocity as a function of temperature gradient. The numerical results and the earlier analytical estimation (dashed line) are reproduced from Ref. 14.

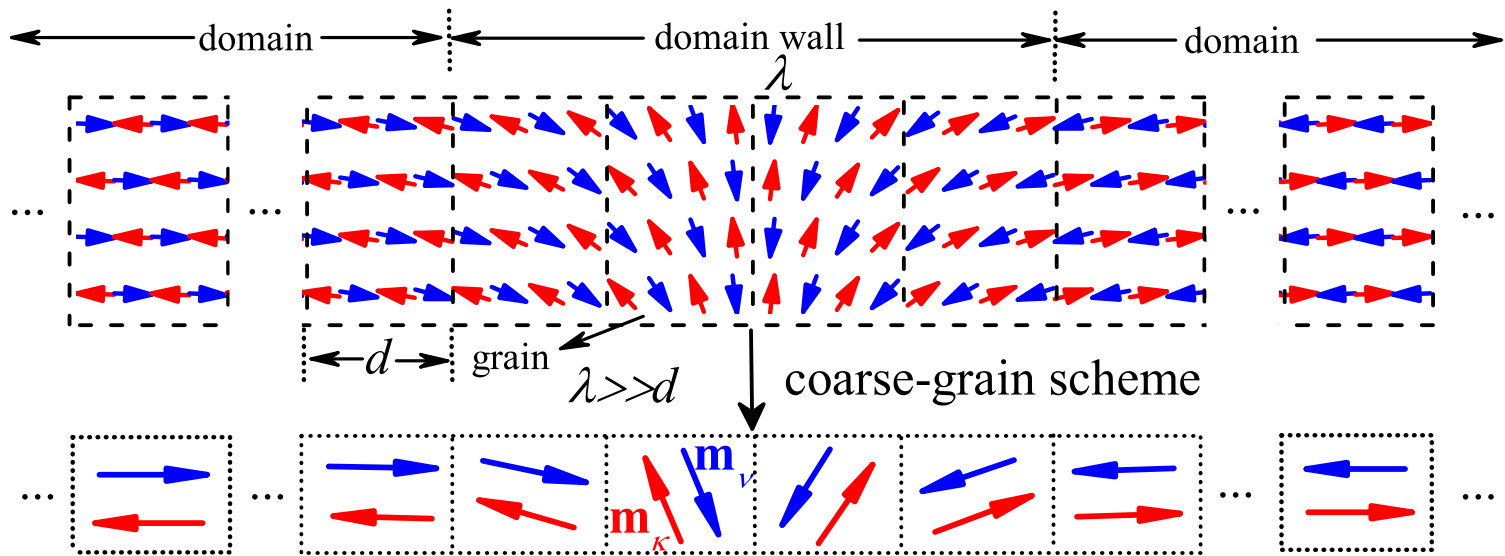

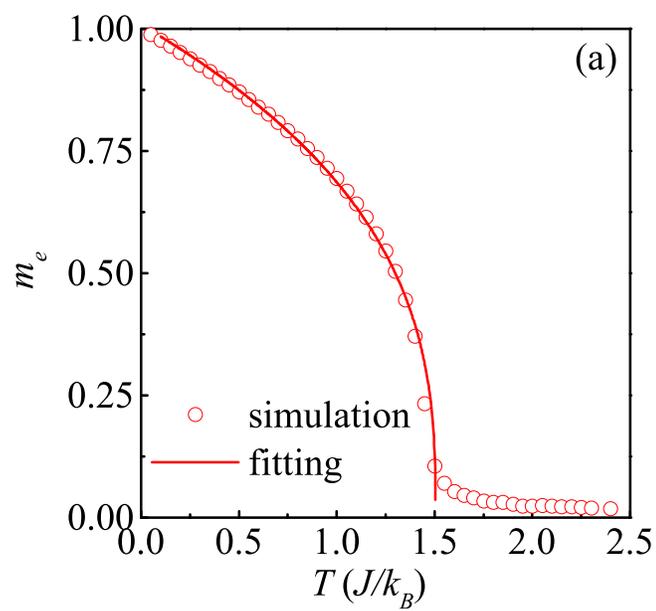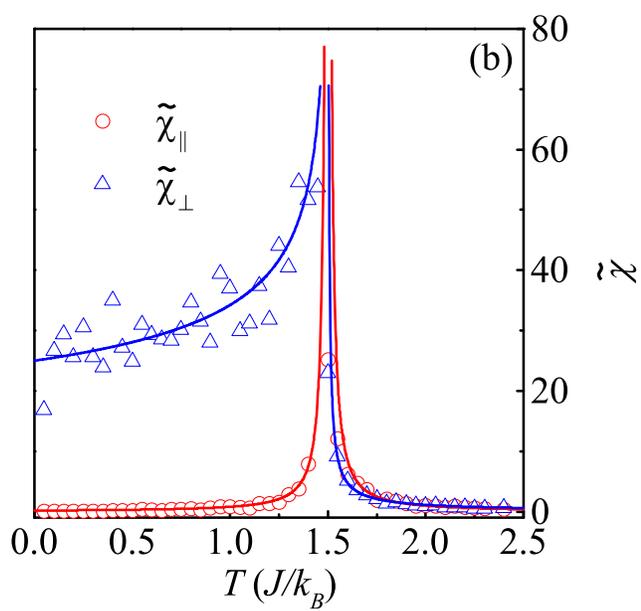

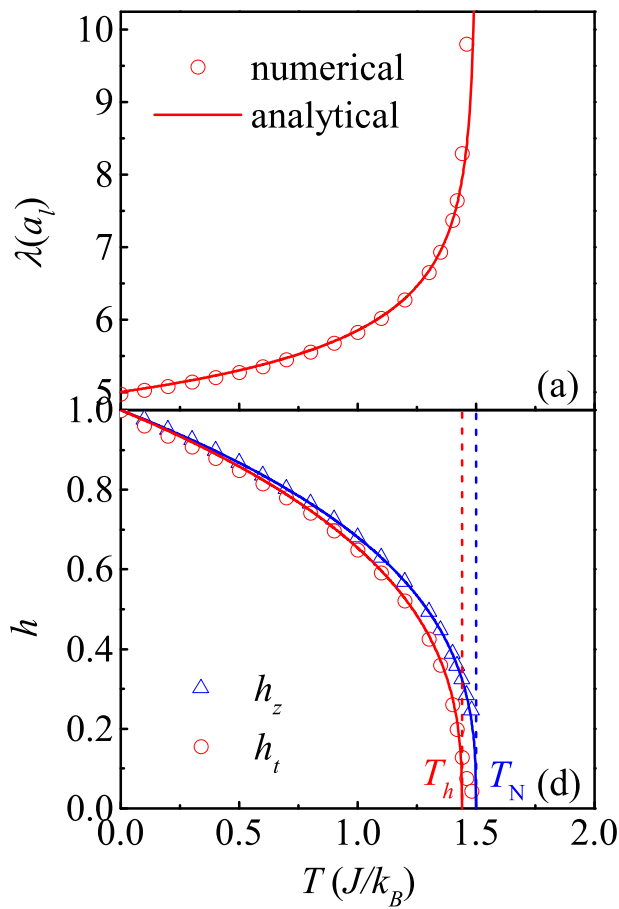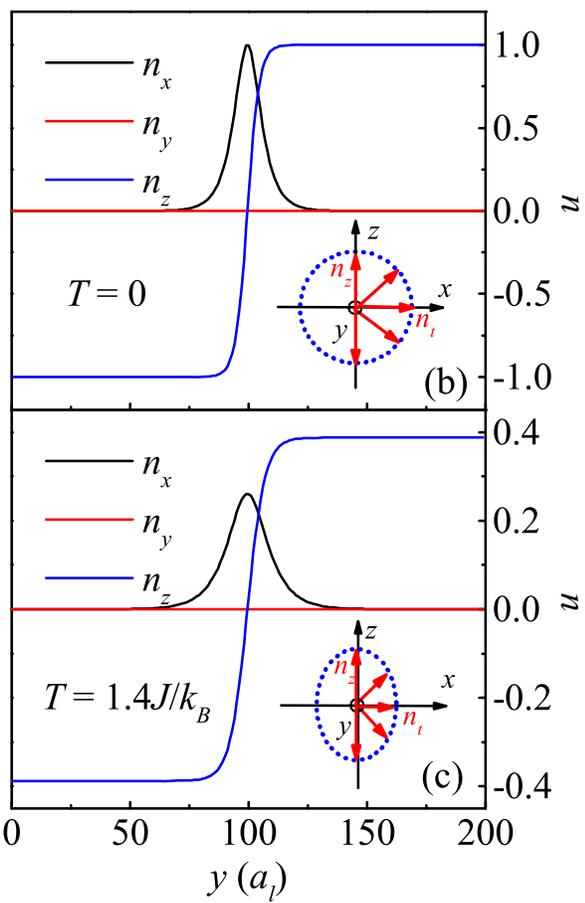

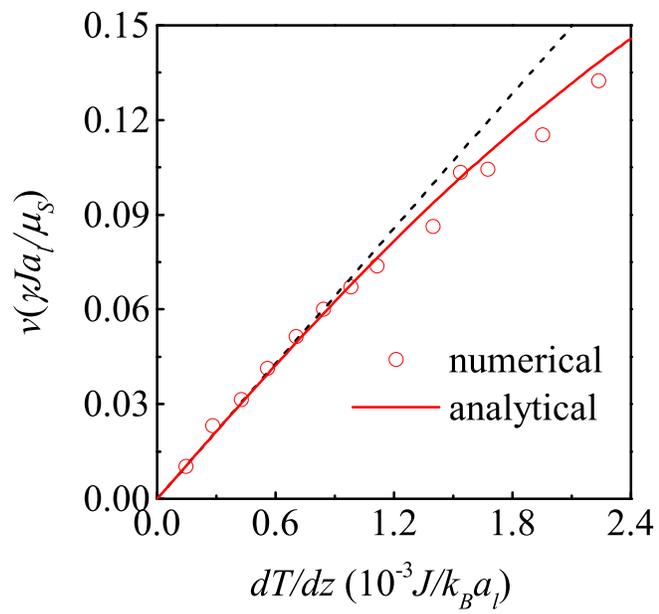

# Supplementary material for "The Landau-Lifshitz-Bloch equation for domain wall motion in antiferromagnets"

Z. Y. Chen[1], Z. R. Yan[1], M. H. Qin[1,*], and J. –M. Liu[1,2]

[1]*Institute for Advanced Materials, South China Academy of Advanced Optoelectronics and Guangdong Provincial Key Laboratory of Quantum Engineering and Quantum Materials, South China Normal University, Guangzhou 510006, China*

[2]*Laboratory of Solid State Microstructures and Innovative Center for Advanced Microstructures, Nanjing University, Nanjing 210093, China*

## A. Continuity equation

After the substitutions and necessary simplification, the LLB equations can be given by

$$\begin{aligned}
\frac{d\mathbf{m}_i}{dt} &= \gamma \left( \mathbf{m}_i \times \mathbf{H}_{m,i} + \mathbf{n}_i \times \mathbf{H}_{n,i} \right) \\
&+ \frac{\gamma \alpha_\parallel}{2m_{e,i}^2} \left[ (\mathbf{m}_i \cdot \mathbf{H}_{m,i})\mathbf{m}_i + (\mathbf{n}_i \cdot \mathbf{H}_{n,i})\mathbf{m}_i + (\mathbf{m}_i \cdot \mathbf{H}_{n,i})\mathbf{n}_i + (\mathbf{n}_i \cdot \mathbf{H}_{m,i})\mathbf{n}_i \right] \\
&- \frac{\gamma \alpha_\perp}{2m_{e,i}^2} \left[ \mathbf{m}_i \times (\mathbf{m}_i \times \mathbf{H}_{m,i}) + \mathbf{m}_i \times (\mathbf{n}_i \times \mathbf{H}_{n,i}) + \mathbf{n}_i \times (\mathbf{m}_i \times \mathbf{H}_{n,i}) + \mathbf{n}_i \times (\mathbf{n}_i \times \mathbf{H}_{m,i}) \right]
\end{aligned} \quad , \quad (1)$$

and

$$\begin{aligned}
\frac{d\mathbf{n}_i}{dt} &= \gamma \left( \mathbf{m}_i \times \mathbf{H}_{n,i} + \mathbf{n}_i \times \mathbf{H}_{m,i} \right) \\
&+ \frac{\gamma \alpha_\parallel}{2m_{e,i}^2} \left[ (\mathbf{m}_i \cdot \mathbf{H}_{n,i})\mathbf{m}_i + (\mathbf{n}_i \cdot \mathbf{H}_{m,i})\mathbf{m}_i + (\mathbf{m}_i \cdot \mathbf{H}_{m,i})\mathbf{n}_i + (\mathbf{n}_i \cdot \mathbf{H}_{n,i})\mathbf{n}_i \right] \\
&- \frac{\gamma \alpha_\perp}{2m_{e,i}^2} \left[ \mathbf{m}_i \times (\mathbf{m}_i \times \mathbf{H}_{n,i}) + \mathbf{m}_i \times (\mathbf{n}_i \times \mathbf{H}_{m,i}) + \mathbf{n}_i \times (\mathbf{m}_i \times \mathbf{H}_{m,i}) + \mathbf{n}_i \times (\mathbf{n}_i \times \mathbf{H}_{n,i}) \right]
\end{aligned} \quad , \quad (2)$$

with the effective fields

---

[*]Authors to whom correspondence should be addressed. Electronic mail: qinmh@scnu.edu.cn

$$\mathbf{H}_{m,i} = \mathbf{H} - \frac{1}{2\tilde{\chi}_\perp}\left(m_{x,i}\mathbf{e}_x + m_{y,i}\mathbf{e}_y\right) - \frac{J_0}{4\mu_S m_{e,i}^2}\mathbf{n}_i \times \left(\mathbf{n}_i \times \mathbf{m}_i\right)$$
$$-\frac{1}{2}\left[\frac{1}{\tilde{\chi}_\parallel}\left(\frac{\mathbf{m}_i^2 + \mathbf{n}_i^2}{4m_{e,i}^2} - 1\right) + \frac{|J_0|}{8\mu_S m_{e,i}^2}\left(\left(\mathbf{m}_i^2 + \mathbf{n}_i^2\right) - \left(\frac{\mathbf{m}_i^2 - \mathbf{n}_i^2}{m_{e,i}}\right)^2\right)\right]\mathbf{n}_i, \quad (3)$$
$$-\frac{\mathbf{m}_i \cdot \mathbf{n}_i}{4m_{e,i}^2}\left(\frac{1}{\tilde{\chi}_\parallel} + \frac{|J_0|}{2\mu_S}\right)\mathbf{m}_i + \frac{A(T)}{d^2 M_S m_{e,i}^2}\sum_j \left(\mathbf{m}_j + \mathbf{m}_i\right)$$

and

$$\mathbf{H}_{n,i} = -\frac{1}{2\tilde{\chi}_\perp}\left(n_{x,i}\mathbf{e}_x + n_{y,i}\mathbf{e}_y\right) - \frac{J_0}{4\mu_S m_{e,i}^2}\mathbf{m}_i \times \left(\mathbf{n}_i \times \mathbf{m}_i\right)$$
$$-\frac{1}{2}\left[\frac{1}{\tilde{\chi}_\parallel}\left(\frac{\mathbf{m}_i^2 + \mathbf{n}_i^2}{4m_{e,i}^2} - 1\right) + \frac{|J_0|}{8\mu_S m_{e,i}^2}\left(\left(\mathbf{m}_i^2 + \mathbf{n}_i^2\right) - \left(\frac{\mathbf{m}_i^2 - \mathbf{n}_i^2}{m_{e,i}}\right)^2\right)\right]\mathbf{m}_i. \quad (4)$$
$$-\frac{\mathbf{m}_i \cdot \mathbf{n}_i}{4m_{e,i}^2}\left(\frac{1}{\tilde{\chi}_\parallel} + \frac{|J_0|}{2\mu_S}\right)\mathbf{n}_i + \frac{A(T)}{d^2 M_S m_{e,i}^2}\sum_j \left(-\mathbf{n}_j + \mathbf{n}_i\right)$$

Based on the continuum approximation, the above equations are respectively transformed into:

$$\frac{d\mathbf{m}}{dt} = \gamma\left(\mathbf{m} \times \mathbf{H}_m + \mathbf{n} \times \mathbf{H}_n\right) - \frac{\alpha_\perp}{2m_e^2}\left(\mathbf{m} \times \frac{d\mathbf{m}}{dt} + \mathbf{n} \times \frac{d\mathbf{n}}{dt}\right)$$
$$+ \frac{\gamma\alpha_\parallel}{2m_e^2}\left[\left(\mathbf{m}\cdot\mathbf{H}_n\right)\mathbf{n} + \left(\mathbf{n}\cdot\mathbf{H}_m\right)\mathbf{n}\right], \quad (5)$$

$$\frac{d\mathbf{n}}{dt} = \gamma\mathbf{n} \times \mathbf{H}_m, \quad (6)$$

$$\mathbf{H}_m = \mathbf{H} - \frac{1}{2\tilde{\chi}_\perp}\left(m_x\mathbf{e}_x + m_y\mathbf{e}_y\right) - \frac{J_0}{4\mu_S m_e^2}\mathbf{n} \times \left(\mathbf{n} \times \mathbf{m}\right)$$
$$-\frac{1}{2}\left[\frac{1}{\tilde{\chi}_\parallel}\left(\frac{\mathbf{m}^2 + \mathbf{n}^2}{4m_e^2} - 1\right) + \frac{|J_0|}{8\mu_S m_e^2}\left(\left(\mathbf{m}^2 + \mathbf{n}^2\right) - \left(\frac{\mathbf{m}^2 - \mathbf{n}^2}{m_e}\right)^2\right)\right]\mathbf{n}, \quad (7)$$
$$-\frac{\mathbf{m}\cdot\mathbf{n}}{4m_e^2}\left(\frac{1}{\tilde{\chi}_\parallel} + \frac{|J_0|}{2\mu_S}\right)\mathbf{m} + \frac{2N_D A(T)}{d^2 M_S m_e^2}\mathbf{m}$$

$$\mathbf{H}_n = -\frac{1}{2\tilde{\chi}_\perp}(n_x \mathbf{e}_x + n_y \mathbf{e}_y) - \frac{J_0}{4\mu_S m_e^2}\mathbf{m}\times(\mathbf{n}\times\mathbf{m})$$
$$-\frac{1}{2}\left[\frac{1}{\tilde{\chi}_\parallel}\left(\frac{\mathbf{m}^2+\mathbf{n}^2}{4m_e^2}-1\right)+\frac{|J_0|}{8\mu_S m_e^2}\left((\mathbf{m}^2+\mathbf{n}^2)-\left(\frac{\mathbf{m}^2-\mathbf{n}^2}{m_e}\right)^2\right)\right]\mathbf{m}. \tag{8}$$
$$-\frac{\mathbf{m}\cdot\mathbf{n}}{4m_e^2}\left(\frac{1}{\tilde{\chi}_\parallel}+\frac{|J_0|}{2\mu_S}\right)\mathbf{n}-\frac{A(T)}{M_S m_e^2}\nabla^2\mathbf{n}$$

## B. The AFM domain wall velocity under a temperature gradient

Substituting Eq. (16) into Eq. (14) in the manuscript and omitting the high-order terms of $n_y$, we obtain

$$\left(\frac{\gamma A}{M_S m_e^2}-A_m v^2\right)(n_z n_x'' - n_x n_z'') - v\frac{\alpha_\perp}{2m_e^2}(n_z n_x' - n_x n_z') = 0, \tag{9}$$

$$\left(\frac{\gamma A}{M_S m_e^2}-A_m v^2\right)(n_x n_x'' + n_z n_z'')n_y + v\frac{\alpha_\perp}{2m_e^2}n_y' - n^2\left(\frac{\gamma A}{M_S m_e^2}-A_m v^2\right)n_y'' = 0, \tag{10}$$

with the prime represents the derivative with respect to $Z$. Transforming the two equations above into the polar coordinates using the following relations:

$$n_x = n\sin(\theta+\psi), \quad n_y = n_y, \quad n_z = n\cos(\theta+\psi), \tag{11}$$

where small quantities $n_y$ and $\psi$ are introduced to describe the slight deformation of the DW, and keeping the leading order of $n_y$ and $\psi$, we obtain

$$(\theta''+\psi'')+B(\theta'+\psi')=0, \tag{12}$$

$$n^2\tilde{A}n_y'' - v\frac{\alpha_\perp}{2m_e^2}n_y' + \tilde{A}\left[(\theta'+\psi')^2 - \frac{n''}{n}\right]n_y = 0, \tag{13}$$

where $B$ is the space-dependent variable

$$B = \frac{2n'}{n} - \frac{\alpha_\perp}{2m_e^2 \tilde{A}} v \tag{14}$$

with

$$\tilde{A} = \frac{\gamma A}{M_s m_e^2} - A_m v^2. \tag{15}$$

We introduce a new variable $\varphi = \theta + \psi$ and substitute it in Eq. (12), and obtain

$$\varphi'' + B\varphi' = 0. \tag{16}$$

Since there is one to one relation between $\theta$ and $Z$, we can substitute $Z$ with $\theta$ and obtain

$$\frac{1}{\sin\theta} \frac{d}{d\theta}\left(\sin\theta \frac{d\varphi}{d\theta}\right) + \frac{\lambda B}{\sin\theta} \frac{d\varphi}{d\theta} = 0. \tag{17}$$

Then transforming Eq. (17) with an introduced variable $X = \cos\theta$, we obtain

$$(1 - X^2)\frac{d^2\varphi}{dX^2} - (2X + \lambda B)\frac{d\varphi}{dX} = 0. \tag{18}$$

Comparing Eq. (18) with the Legendre's polynomials:

$$(1 - x^2)\frac{d^2 y}{dx^2} - 2x\frac{dy}{dx} + l(l+1)y = 0, \tag{19}$$

with an integer $l$, we know that Eq. (18) has a steady solution only for $B = 0$. A divergence of the solution is obtained for a nonzero $B$, which is inconsistent with the consideration of the steady AFM DW motion. Consequently, we obtain a quadratic equation of the DW velocity:

$$v = \alpha_1\left(1 - \alpha_2 v^2\right). \tag{20}$$

Solving this equation and disregarding the unphysical solution, we obtain

$$v = \frac{-\frac{1}{\alpha_1} + \sqrt{\frac{1}{\alpha_1^2} + 4\alpha_2}}{2\alpha_2}. \tag{21}$$

Here, the *T*-dependent variables $\alpha_1$ and $\alpha_2$ read

$$\alpha_1 = -\frac{2\gamma A(0)}{\alpha_\perp M_S T_N} \nabla T, \quad \alpha_2 = \frac{1}{2N_m}\left(\frac{\mu_S}{\gamma a_l J m_e}\right)^2. \tag{22}$$

For small *T* gradients, Eq. (21) can be approximated to

$$v = \alpha_1 = -\frac{2\gamma A(0)}{\alpha_\perp M_S T_N} \nabla T, \tag{23}$$

which well explains the agreement of the equation with numerical results reported in earlier work.

### *C. The derivation and modification of group velocity of spin waves*

In this section, following Ref. 20, we derive the group velocity of spin waves at finite *T* by solving Eq. (14) in the manuscript and neglecting the dissipation,

$$A_m \mathbf{n} \times \ddot{\mathbf{n}} \times \mathbf{n} = \mathbf{n} \times \left(-\frac{\gamma}{2\tilde{\chi}_\perp} n_z \mathbf{e}_z + \frac{\gamma A(T)}{M_S m_e^2} \nabla^2 \mathbf{n}\right) \times \mathbf{n}. \tag{24}$$

The low excitation of spin waves can be described by the following equation,

$$\mathbf{n}(\mathbf{r},t) = n\mathbf{e}_z + \delta \mathbf{n}_\perp e^{i(\omega t - \mathbf{k}\cdot\mathbf{r})}, \tag{25}$$

where $\delta \mathbf{n}_\perp$ is the small space-and-time-independent quantity, $\omega$ and $\mathbf{k}$ are the angular frequency and wave vector of spin waves, respectively. Substitute Eq. (25) into Eq. (24), and we obtain the dispersion relation,

$$\omega^2 = \frac{1}{A_m}\left(\frac{\gamma A}{M_S m_e^2}k^2 - \frac{\gamma}{2\tilde{\chi}_\perp}\right), \tag{26}$$

In the isotropic limit, we get a more compact expression,

$$\omega = c(T)k, \tag{27}$$

where $c(T)$ is the group velocity of spin waves at finite $T$,

$$c(T) = \sqrt{\frac{\gamma A}{A_m M_S m_e^2}} = \frac{\gamma a_l J \sqrt{2N_D}}{\mu_S} m_e. \tag{28}$$

At zero $T$, $c(0)$ is identical to the preceding one derived from the LLG equation.

It is well known that the DW velocity is limited by the group velocity of spin waves. Therefore, the maximum of the DW velocity $v_{max}$ could be obtained under a critical $T$ gradient which can be estimated approximately by:

$$\frac{2\gamma|A(0)|}{\alpha_\perp M_S T_N}\nabla T = c(L \cdot \nabla T), \tag{29}$$

where $L$ is the length of the AFM nanowire used in the LLG simulation. Based on this equation, the group velocity of spin wave at finite $T$ is estimated and used to modify the value of $\alpha_2$. Thus, the DW velocity in Eq. (21) could be reasonably modified to describe the reality more precisely.

### D. The effect of magnetic anisotropy gradient induced by temperature gradient

To figure out the effect of the magnetic anisotropy, we transform Eq. (14) in the manuscript into polar coordinates with the assumption that the DW lies in the $x$-$z$ plane while moves along the nanowire, and obtain

$$2\tilde{A}\left(\frac{n'\sin\theta}{n\lambda} + \frac{\sin\theta\cos^3\theta}{\lambda^2}\right) - \frac{\alpha_\perp \sin\theta}{2m_e^2\lambda}v + \frac{\gamma}{2\tilde{\chi}_\perp}\sin\theta\cos\theta = 0, \tag{30}$$

where $\theta$ is the angle between the staggered magnetization and $z$-axis, and the derivatives of $\theta$ with respect to time have been substituted with Eq. (16) in the manuscript. After the integration over the whole space, we get the following equation:

$$2\tilde{A}\frac{n'\pi}{n\lambda} - \frac{\alpha_\perp \pi}{2m_e^2 \lambda}v + K_T = 0, \tag{31}$$

where $K_T$ is related to the effect of the magnetic anisotropy gradient induced by the temperature gradient,

$$K_T = \int dx \frac{\gamma}{2\tilde{\chi}_\perp} \sin\theta \cos\theta. \tag{32}$$

We solve Eq. (31) and obtain

$$v = \frac{-\frac{1}{\alpha_{K,1}} + \sqrt{\frac{1}{\alpha_{K,1}^2} + 4\alpha_{K,2}}}{2\alpha_{K,2}}, \tag{33}$$

with

$$\alpha_{K,1} = \frac{2m_e^2}{\alpha_\perp}\left(\frac{2\gamma n'A(T)}{nM_s m_e^2} + \frac{\lambda}{\pi}K_T\right), \alpha_{K,2} = A_m\left(\frac{\gamma A(T)}{M_s m_e^2} + \frac{\lambda n}{2\pi n'}K_T\right)^{-1}. \tag{34}$$

For small temperature gradients, the difference of the magnetic anisotropy along the nanowire could be neglected, and $K_T = 0$ is obtained. Thus, Eq. (33) could be approximated replaced by Eq. (21).